\documentstyle[12pt]{article}
\begin{document}
\newcommand{\preprint}[1]{\begin{table}[t]  
           \begin{flushright}               
           \begin{large}{#1}\end{large}     
           \end{flushright}                 
           \end{table}}                     

\baselineskip 18pt
\preprint{TAUP-2216-94}

\newcommand{\be}{\begin{equation}}
\newcommand{\ee}{\end{equation}}
\title{Some Interesting
Properties of Field theories with an Infinite Number of Fields}
\author{N.Itzhaki\\Raymond
and Beverly Sackler Faculty of Exact Sciences\\School
 of Physics and Astronomy\\Tel Aviv University, Ramat Aviv, 69978, Israel}
\maketitle
\begin{abstract}
We give an indication that gravity coupled to an infinite number of fields
might be a renormalizable theory.
A toy model with an infinite number of interacting fermions in four-dimentional
 space-time
 is analyzed.
The model is finite at any order in perturbation theory.
However, perturbation theory is valid only for external momenta smaller than
$\lambda ^{-\frac{1}{2}}$ , where $\lambda$ is the coupling constant.
\end{abstract}
\newpage
\section{Motivation and Introduction}
Since $G$, Newton constant has dimension $[mass]^{-2}$, it has been suspected
for a long time that the Einstein-Hilbert lagrangian describes
a  non-renormalizable theory.
In \cite{gr,ve}'t Hooft and Veltman used dimensional
regularization and the background
field method  to show that gravitation coupled to a scalar field is indeed
non-renormalazibale.

In particular 't Hooft and Veltman showed that  the lagrangian
\be {\cal L}=\sqrt{g}(-\frac{1}{2}\partial_{\mu} \psi_{i} g^{\mu \nu}
\partial_{\nu}
\psi_{i} + \frac{1}{2}\psi_{i} M_{ij} \psi_{j}).\ee
which describes scalar fields in an external  gravitational field generates the
 one loop counter-Lagrangian
\be \Delta {\cal L}=\frac{\sqrt{g}}{\epsilon}
Tr[\frac{1}{4}(M_{ij}-\frac{1}{6}R)^{2}
+\frac{\delta _{ij}}{120}(R_{\mu\nu}R^{\mu\nu}-\frac{1}{3}R^{2})]\ee
where the trace is taken over $i$ and $j$.
The first term in $\Delta {\cal L}$ can be absorbed by a redefinition of
 parameters in the lagrangian: $ M\longrightarrow M+\frac{1}{6}R $,
which leads to the improved energy-momenton tensor \cite{cl}.
However the second term is  problematic  since it
can not be absorbed by a redefinition
of parameters in ${\cal L}$.
Since the sighn before $R_{\mu\nu}R^{\mu\nu}$ does not depend
on the type of quantized
field as long as they are real,
 the theory is nonrenormalizable for a finite number of particles.
Let us consider the case of infinite number of fields.

In particular let us consider the following theory.
\be {\cal L}= \sum_{i=0}^{\infty}{\cal L}_{i}\ee
where
\be {\cal L}_{i}=\sqrt{g} (-\frac{1}{2}\partial_{\mu}\overline{\psi}_{i}
g^{\mu\nu}
\partial_{\nu}\psi_{i} +\frac{1}{2}m_{i}^{2}\mid \psi_{i}\mid^{2}
)\ee
 $\psi _{i}$ is a scalar field with $c_{i}$ components and with mass $m_{i}$
{}.
By definition $c_{0}=1$ and $c_{i}=12i$ for $i>0$.
Therefore the lagrangian has a symmetry $U(1)\times
\prod_{i=1}^{\infty}SU(12i)$.
Following [1] we calculate $\Delta {\cal L}$ to obtain
\be \Delta {\cal L}^{p}=\frac{\sqrt{g}}{120\epsilon}
(R_{\mu\nu}R^{\mu\nu}-\frac{1}{3}R^{2})(1+12\sum_{i=1}^{\infty}i)\ee
 where $\Delta {\cal L}^{p}$ is the problematic term in $\Delta {\cal L}$
(meaning the term which can not be absorbed by a
redefinition of parameters in ${\cal L}$).
Although we use dimensional regularization
the theory is obviously not finite even at
 $n\neq 4$, since the sum diverges.

In order to regularize the sum we  consider the same theory but with
\[ c_{i} = \left\{ \begin{array}{ll}
                     1   & \mbox{ i=0} \\
                    12i^{-s}   & \mbox{$i>0$}
                 \end{array}
          \right.\]
this theory yields
\be \Delta {\cal
 L}^{p}=\frac{1}{\epsilon}R_{\mu\nu}R^{\mu\nu}(1+12\sum_{i=0}^{\infty}i^{-s}).
\ee
 For $s>1$ the sum and $ \Delta {\cal L}$ are  finite, hence the theory
 is regularized and well defined.
 For  $s<1$ the theory is redefined as the analytic continuation of the
$\zeta$-
 function.
\be \Delta {\cal L}^{p}=\frac{1}{\epsilon}R_{\mu\nu}R^{\mu\nu}(1+12\zeta
(s))\ee
Since $\zeta (-1)=-\frac{1}{12}$ we get for $s=-1$,
\be \Delta {\cal L}^{p}=0\ee

The natural questions which arise now are under which conditions does
$\Delta {\cal L}^{p}$  vanish when
 gravitation is also quantized   and whether it occurs at any order of
 perturbation theory.
Due to the complexity     of gravitation all we can  say at
 the moment is that it seems that in order to obtain $\Delta {\cal L}^{p}=0$
even at the one loop order, one must introduce a tower of massive spin two
fields . In doing so one must consider the inconsistency problems \cite{de}
that arise in coupling a massive spin two particles to gravity.
This inconsistency is absence in Kaluza-Klein theory \cite{du}.
Therefore it seems that the generalization of the theory to a case in which
gravitation is also quantized   involves an infinite
dimensional Kaluza-Klein theory .
In this paper we consider a simpler non-renormalizable theory
with a coupling constant whose  dimension is $[mass]^{-2}$- four fermi theory.
The outline of the paper is as follows.
In sec.2 we present the model to be discussed.
In sec.3 we prove that the model is finite at any order of perturbation theory.
In sec.4 we show that the Froissart bound is violated since for large external
 momenta perturbation theory is not valid.

\section{The Model}
It is a well known fact that a four Fermi theory which is described by the
 lagrangian
\be{\cal
 L}_{fer}=\overline{\Psi}(i\partial\hspace{-2mm}/ -m)
\Psi+\frac{1}{2}\lambda(\overline{\Psi}\Psi
 )^{2}\ee
is a non-renormalizable theory since it yields
infinities that can not be  absorbed
by a redefinition of parameters in the lagrangian.
That can be seen easily from power counting arguments which give
\be D=I-(E-4)\ee
I is the number of internal legs , E is the number of external legs and $D$
is the degree of divergence.
The model that we shall consider here will contain an infinite number of
fermion
 fields with
quartic interactions in four-dimensional space-time .
The  model describes the same semi-classical low energy physics as ${\cal
 L}_{fer}$.
The model is described by the lagrangian
\be {\cal L}_{2}=\sum_{i=0}^{\infty} {\cal L}_{i}^{free} +{\cal L}_{int
}\ee
where
\be {\cal L}_{i}^{free}=\overline{\Psi}_{i}
(i\partial\hspace{-2mm}/ -m_{i})\Psi_{i}\ee
where $\Psi_{i}$ is a $c_{i}$ component, fermionic field.

The "electrons" are described by $\Psi_{o}$ therefore $c_{0}=1$ and
 $m_{0}=m_{e}$.
In order that ${\cal L}_{2}$ describes the same classical low energy
 physics as ${\cal L}_{fer}$ we must impose
$m_{i}\gg m_{e} $ $ \forall i>0$.
Furthermore, $m_{i}$ and $c_{i}$ satisfy the following conditions:
 \be (\sum_{i=0}^{\infty} c_{i}m^{a}_{i})_{reg}=0 ,\;\;a=0,1,2,3 \ee
and their asymptotic behavior (refers to i) is

\be m_{i}=Me^{i}\;\; ,\;\; c_{i}=ci.\ee
Those conditions are fulfilled by

\[  c_{i}=\left\{ \begin{array}{ll}
                     1      & \mbox{if $i=0$} \\
                     12i  & \mbox{otherwise}
                   \end{array}
          \right. \]
\be m_{i}=\left\{ \begin{array}{ll}
                     m_{e}   & \mbox{if i=0} \\
                     Me^{i} & \mbox{$i>3$}
                  \end{array}
          \right.\ee
$m_{1} , m_{2} , m_{3}$ are defined  by eq.(12) for $a=1,2,3$.

\begin{equation}{\cal L}_{int} =\lambda((\sum_{i=0}^{\infty}\overline{\Psi}_{i}
)(\sum_{i=0}^{\infty}\Psi_{i}))^{2}\end{equation}
 Note, that ${\cal L}_{int}$ breaks the
$U(1)\times\prod_{i=1}^{\infty}SU(12i)$
 symmetry of ${\cal L}_{free}$ and leaves only the $U(1)$ symmetry
$ \Psi_{i}\rightarrow e^{i\alpha}\Psi_{i} ,\overline{\Psi}_{i}\rightarrow
 e^{-i\alpha}\overline{\Psi}_{i} $
which generates the conservation of the total fermion number.

\section{Finiteness}

In this section we use dimensional regularization and we prove that at any
order
 of perturbation theory the Green functions are finite.
 In units where $M=\lambda =1$ $\hbar$ is a small dimensionless parameter (when
 $\lambda M^{2}\ll 1$ in  units where $\hbar =c=1$),and can be considered as
 expansion
parameter of  perturbation theory.

First let us show that the poles for $n=4$  vanish.
We will treat one and two loop diagrams.
The generalization to higher loops is obvious.
Consider first one loop diagrams.
The residues of eventual poles for $n=4$ are finite polynomials in the momenta
 and masses [2].
Since by definition , the coupling constant is a constant function  of the
 fields,
$\lambda$,
we get
\begin{equation}{ G}(p_{1},..p_{E})=\frac{1}{\epsilon}
 \sum_{a_{
j}}\prod_{j=1}^{I}(\sum_{i=0}^{\infty}c_{i}m_{i}^{a_{j}})g(a_{j},n,p_{1},..p_{E
})+{\cal O}(\epsilon ^{0})\end{equation}
In order to simplify the discussion we do not include Feynman parameters which
 are irrelevant for the discussion.
As in the first section we see that although we use dimensional regularization
 the Green functions are not finite even for $n\neq 4$, since the
sum diverges and must  be regularized.
In order to regularized the theory we treat the same theory but with
\be c_{i}=\left\{ \begin{array}{ll}
                     1   & \mbox{if i=0} \\
                    12i^{-s}  & \mbox{otherwise}
                  \end{array}
          \right.\ee
\be m_{i}=\left\{ \begin{array}{ll}
                     m_{e}   & \mbox{if i=0} \\
                     Mq^{i} & \mbox{$i>3$}
                  \end{array}
          \right.\ee
This theory is identical to the original theory at $s=-1$ and $q=e$.
Again $m_{1} , m_{2} , m_{3}$ are defined  by eq.(12) for $a=1,2,3$.
It is easily  seen that the residues for $n=4$ are finite for $s>1$ and $\mid
 q\mid<1$.
The theory is redefined for $s=-1 , q=e$ as the analytic continuation of the
 well defined theory ($s>1 , \mid q\mid<1$).
Dimensional analysis implies that the maximal order         of the polynomial
is
 $3$.
Therefore we should consider only
 \be (\sum_{i=0}^{\infty} c_{i}m^{a}_{i})_{reg}=0 ,\;\;a=0,1,2,3 \ee
In sec.1 we show that
\be (\sum_{i=0}^{\infty} c_{i})_{reg}=0.\ee
Let us treat now \be (\sum_{i=0}^{\infty} c_{i}m_{i})_{reg}.\ee
If $m_{0}=m_{e}$ and $m_{i}=Mq^{i}$ for $i>0$.
then
\be
 \sum_{i=0}^{\infty}c_{i}m_{i}=m_{e}-12M\frac{q}{(1-q)^{2}},\ee
for $ \mid q\mid<1$ .
Thus the analytic continuation for $q=e$ gives
\be (\sum_{i=0}^{\infty} c_{i}m^{a}_{1})_{reg}=m_{e}-12\frac{e}{(1-e)^{2}}M\ee
therefore when
\be m_{i}=\left\{ \begin{array}{ll}
                     m_{e}  & \mbox{ i=0} \\
                     Me(1+\frac{e}{(1-e)^{2}})-\frac{1}{12}m_{e} & \mbox{i=1}
\\
                     Me^{i}   & \mbox{$i>1$}
                   \end{array}
          \right.\ee
we get
\be (\sum_{i=0}^{\infty} c_{i}m^{a}_{1})_{reg}=0\ee
In principle we can  also arrange
 \be (\sum_{i=0}^{\infty} c_{i}m^{a}_{i})_{reg}=0 ,\;\;a=2,3 \ee
 by changing $m_{2} , m_{3}$.
Thus the residues at the one loop order  vanish.
Next we turn to two loops.

{\it Definition}:[2] A harmless pole is a
pole whose  residue is a polynomial of finite order in the external momenta and
 masses.

{\it Theorem 1} Two-loop diagrams  contain only harmless poles.

{\it Proof}: In \cite{th} it has been shown that   the residue of a two loops
 diagram including the contributions from counterterms of one loop subdiagrams
 contains only a harmless poles.
It was shown above that there are no counterterms in the one
loop  diagrams, thus the residue of any two loops diagrams  contains only
 harmless poles.

{\it Theorem 2} In two loops diagrams the residues  vanish.

{\it Proof} From theorem 1 it is clear that for any Feynman diagram
\begin{equation}{ G}(p_{1},..p_{E})=\sum_{n=1}^{L}\frac{1}{\epsilon^{n}}
 \sum_{a_{
j}}\prod_{j=1}^{I}(\sum_{i=0}^{\infty}c_{i}m_{i}^{a_{j}})g(a_{j},n,p_{1},..p_{E
})+{\cal O}(\epsilon ^{0})\end{equation}
where L is the number of loops.
{}From dimensional analysis and power counting arguments one gets
\begin{equation}\sum_{j=1}^{I}a_{j}\leq D \leq
 3I \end{equation}
Therefore, at least for one of the internal legs $a_{j}\leq 3$
and the residues vanish.
Let us remark that the generalization to higher loops is obvious since eq.(29)
does not depend on the number of  loops.
Therefore in this theory the poles at $n=4$ do not generate counterterms as
fields theories with a finite number of fields do.
Still,  we are left with expressions of the form
\be (\sum_{i_{1}}..\sum_{i_{I}} c_{i_{1}}m_{i_{1}}^{a_{i_{1}}}..c_{i_{I}}
 m_{i_{I}}^{a_{i_{I}}}\ln{P(m_{i_{1}},..m_{i_{I}},P_{ext}}))_{reg}\ee
which might diverge (again we do not include Feynman parameters).
Since for $i>4$ , $c_{1}=12i$ and $m_{i}=Me^{i}$,
eq.(30) has the following schematic  form
\be \sum_{i=0}^{\infty}i^{d}e^{ic}\;\;\; ,\;{\mbox where}\;\; d>0\ee
which is well defined for $c<0$, and
 can be redefined as the analytic continuation of
\be P(\frac{\partial}{\partial q})\frac{1}{1-q}\mid_{q=e^{c}}\ee
when $c>0$,
where $P$ is a finite polynomial of $\frac{\partial}{\partial q}$
 of order $d$.
When $c=0$ it can be redefined using $\zeta $-function,
 since $d>0$.
Notice further that eq.(12) is closely related
to Pauli-Villars regularization equations.
However, in Pauli-Villars one use a finite number of fields,
hence in order to satisfy
\be (\sum c_{i})=0\ee
part of the $c_{i}$ must be positive (real fields) and part of the $c_{i}$ must
 be negative (ghost fields).
 Thus the masses of the ghosts must go to infinity, otherwise unitarity will be
explicitly violated          .
In our theory we satisfy
          \be (\sum c_{i})_{reg}=0 \ee
using an infinite number of real fields, and limit
 $M\longrightarrow \infty$ need not be taken.
\section{Unitarity and Scaling Behaviour}
The proof of the finiteness of the theory was based
on a redefinition of divergent  sums using
 analytic continuation.
This method might be dangerous when one considers unitarity, since the
 regularized value of a sum of positive elements might be negative,
and if for example the elements are $\sigma _{1,2\longrightarrow a,b}(>0)$ then
$\sigma_{tot}\leq \sum_{a,b}\sigma _{1,2\longrightarrow a,b} $
might be negative.
In that case the theory is simply nonsense.
In other words we must prove that the imaginary part of the amplitude is
 positive.
Let us consider for example, the diagram of Figure~1;  we must show that
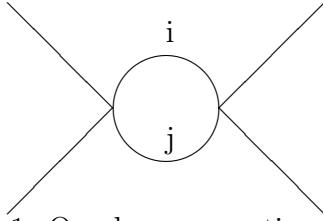
\begin{figure}
\begin{picture}(300,50)(0,0)
\put(150,25){\circle{40}}
\put(150,10){j}
\put(150,50){i}
\put(170,25){\line(1,1){40}}
\put(170,25){\line(1,-1){40}}
\put(90,65){\line(1,-1){40}}
\put(90,-15){\line(1,1){40}}
\end{picture}
\caption{One loop correction to the coupling constant}
\end{figure}
\begin{eqnarray} Im[-\sum_{i,j}c_{i}c_{j}\int_{0}^{1}dx
(\log(\frac{xm_{i}^{2}+(1-x)
m_{j}^{2}+p^{2}x(1-x)}{\mu^{2}})\\ \nonumber
((xm_{i}^{2}+(1-x)m_{j}^{2}+p^{2}x(1-x))]_{reg}>0,\end{eqnarray}
otherwise the total cross section will be negative.
It is obvious that $ -p^{2}x(1-x)>xm_{i}+(1-x)m_{j}$ only when
$ -p^{2}>m_{i}^{2}+m_{j}^{2}.$
Therefore, if $\lim_{i \rightarrow \infty} m_{i} = \infty$ (as in our case)
 then for any finite $p$ the sum is taken
over a finite number of positive elements
 and is  positive,
caussing no violation of unitarity .
However, it is easy to see that as in the
usual four Fermi interaction  the Froissart
 bound is violated.
This is  not surprising since in  units
where $\hbar =c=1$  $\lambda$ has dimension
of  $[mass]^{-2}$,
hence when
$p^{2}>\frac{1}{\lambda}$   the effective expansion parameter is larger
 than $1$ and
the perturbation expansion is meaningless.
Another way to illustrate the problem is to analyze the behavior of the theory
 under a scale transformation.

Since the theory is finite,
\be\beta =\gamma =\gamma_{m} =0,\ee
and therefore the scaling equation is trivial:
\be\Gamma_{n}(sp,m,\lambda)=s^{d_{n}}\Gamma_{n}(p,m(s),\lambda(s))\ee
where
\be s\frac{\partial m}{\partial s}=-m\ee
\be s\frac{\partial \lambda}{\partial s}=2\lambda.\ee
The solution is
\be\Gamma_{n}(sp,m,\lambda)=s^{d_{n}}\Gamma_{n}(p,\frac{m}{s},\lambda s^{2})\ee
In particular we see that
\be \lambda\rightarrow \lambda s^{2}\ee
Let us remark that if instead of the four constraints of eq.(12) one
constructs a theory with only one condition
\be (\sum_{i=0}^{\infty} c_{i})_{reg}=0\ee
then the theory will not be finite , but  the counterterms that one need to add
 are of the form of terms in the original lagrangian \footnote{It is due
to the fact that $D=I-(E-4)$, hence for $E>4$ $D<I$.}.
In that case we get
\be s\frac{\partial \lambda}{\partial s}=2\lambda+{\cal O}(\hbar).\ee
Since $\hbar \ll 1$ (otherwise perturbation theory is meaningless even at low
 energy) the
second term can not change the asymptotic behavior and the fact that for
 $p^{2}> \frac{1}{\lambda}$ the behavior  of the model is not described by
 perturbation theory.

\section{Discussion}
It is usually said that field
theories with a coupling constant whose dimension is
$[mass]^{-n}$  $n>0$ are nonrenormalizable
theories since their perturbation expantion yields infinities
which can not be absorved by a redefinition of parameters in the lagrangian.
We have been able to find  a
theory with coupling constant of dimension $[mass]^{-2}$ which is finite
at any order in perturbation theory.
However,  due to the
dimension of the coupling constant the problems reappear when one considers
Green functions with external momenta larger than $\lambda ^{-\frac{1}{2}}$.
At those momenta perturbation expansion is no longer valid,
which means that the
small distance
behaviour of the theory is not described by the usual perturbation expansion  .
The same problem appears in field theories with a dimensless coupling constant
with a positive $\beta $ function.
However there is a major difference between the two cases:
in our toy model perturbation theory is not valid for
$p>\lambda ^{-\frac{1}{2}}$ while in Q.E.D
for instance the breakdown occurs only at $M e^{\frac{1}{\alpha}}$.
Moreover since in our toy model there are  particles with mass larger than
$\lambda ^{-\frac{1}{2}}$  it is not
clear whether they are the right excitations of the theory (if there are any).
Note that those  excitations are the key in the proof of  finiteness,
hence it is not clear whether one can consider this theory as a physical
theory at any scale.
 Since the region in which
the perturbation theory is trustworthy is that of low momentum transfer it
 is  essential to find a nonperturbative alternative description
at high energy. One
possible way of doing so is to investigate the model on a lattice.
The basic requirement imposed on the lattice version of a continuum theory
is that it reproduce the continuum limit as the lattice spacing $a$ tends
to zero.
But this can not be the case in our model since for any finite $a$ there are
an infinite  number of fields with masses larger
than $\frac{1}{a} $ while for $a=0$ $\frac{1}{m_{i}}>a$ $\forall i$.

For those reasons it is an open question whether ${\cal L}_{2}$ describes
 a consistent theory.
\vspace{1.5cm}

I am grateful to  Prof. A.Casher for pointing  out the ultraviolet problem of
the theory and for helpful discussions.
I would also like to thank  Prof. Y.Aharonov
and Prof. S.Yankielowicz for useful discussions.

\end{document}